\documentclass[12pt]{article}
\usepackage{amssymb,amsmath}
\usepackage[colorlinks=false,
      linkcolor=darkblue,  
      urlcolor=blue,    
      filecolor=blue,     
      citecolor=red,
linktocpage=true,
      pdfstartview=FitV,
      bookmarksopen=true]{hyperref}
\usepackage{epsfig}
\usepackage{epstopdf}
\usepackage{latexsym}
\usepackage{graphicx}
\usepackage{subfigure}
\pdfoutput=1

\numberwithin{equation}{section}

\def\IR{\mathbb{R}}

\def\ZZ{\mathbb{Z}}

\def\cA{{\cal A}}

\newcommand{\ac}{\acute}

\newcommand{\si}{\sigma}
\newcommand{\lm}{\lambda}
\newcommand{\ka}{\kappa}
\newcommand{\al}{\alpha}

\newcommand{\omg}{\omega}

\newcommand{\bq}{\begin{equation}}
\newcommand{\eq}{\end{equation}}
\newcommand{\bs}{\begin{split}}
\newcommand{\es}{\end{split}}
\newcommand{\ra}{\rightarrow}

\begin{document}  

\begin{titlepage}
 
\bigskip
\bigskip
\bigskip
\bigskip
\begin{center} 
{\Large \bf   Corrugated Multi-Supersheets}

\bigskip
\bigskip 

{\bf 
Orestis Vasilakis \\
 }
\bigskip
\bigskip
Department of Physics and Astronomy \\
University of Southern California \\
Los Angeles, CA 90089, USA  \\
\bigskip
~vasilaki@usc.edu  \\
\end{center}

\begin{abstract}

\noindent 
We explore the multi-superthreads and supersheets solutions of six-dimensional $\mathcal{N}=1$ supergravity coupled to a tensor multiplet. The solutions carry $D1$-$D5$-$P$ charges, but no Kaluza-Klein monopole. We lay down the formalism to construct multiple supersheets with arbitrary and independent profiles. The solution is by construction free of Dirac strings in contrast to the five-dimensional construction where one has to separately solve integrability conditions. We explore this formalism to construct supersheets that fluctuate in both directions allowing a more general choice of profiles. These new solutions are genuinely six-dimensional, singular, fluctuating BPS solutions and by analyzing them we expect to learn more about the conjectured superstrata. We also derive the conditions under which different supersheets can touch, or even intersect through each other.
\end{abstract}

\end{titlepage}


\tableofcontents

\section{Introduction}
A lot of work has recently been done in exploring BPS solutions of $\mathcal{N}=1$ six-dimensional supergravity coupled to an anti-symmetric tensor multiplet. In \cite{Bena:2011dd}, by exploring the results of \cite{Gutowski:2003rg,Cariglia:2004kk}, it was shown that the BPS equations are linear. Linearity \cite{Bena:2004de} of the BPS equations in five dimensions \cite{Gauntlett:2002nw} has allowed the construction and classification of many three-charge solutions both supersymmetric \cite{Bena:2007kg,Berglund:2005vb}, and non-supersymmetric ones  \cite{Bena:2009fi, Goldstein:2008fq, Bena:2009ev, Bena:2009en, Vasilakis:2011ki, Bobev:2011kk, Bobev:2009kn, Dall'Agata:2010dy, Bossard:2011kz}.  Consequently it is expected that the linearity of BPS equations in six dimensions will also allow such a classification for solutions that cannot be reduced to five dimensions.

Interest in solutions of six-dimensional supergravity was also raised by the conjectured superstrata, novel microstate geometries with three electric and three dipole magnetic charges that depend non-trivially in the compactification direction and thus are genuinely six-dimensional.  Finding smooth supergravity geometries with the same asymptotic charges as black holes is of central importance in the fuzzball proposal \cite{Lunin:2001jy, Lunin:2002iz, Mathur:2005zp, Chowdhury:2010ct, Mathur:2008nj, Balasubramanian:2008da, Skenderis:2008qn}, which tries to explain the microscopic structure of black holes. Many of these microstates are expected to be exotic brane configurations which in general are non-geometric  \cite{deBoer:2010ud, deBoer:2012ma}. However, in the D1-D5-P duality frame one of these configurations can admit a geometric description and thus it may also be smooth. It is expected that the superstratum, because of its additional fluctuation modes in the compactification direction, will contribute significantly to the microstates of three-charge geometries at a semiclassical level. 

The existence of the superstratum was argued based on supersymmetry arguments in \cite{Bena:2011uw}. However its full description in terms of an exact supergravity solution is yet to be found. Some first steps towards that direction \footnote{A different perturbative approach towards the superstratum based on string amplitudes has recently been given in \cite{Giusto:2012jx}. } along with the explicit construction of two and three charge solutions in six dimensions were done in \cite{Niehoff:2012wu, Bena:2011dd}. In \cite{Bena:2011dd}  a one-dimensional D1-D5-P geometry with two dipole charges and arbitrary profile was constructed and named superthread. The linearity of the BPS equations was further explored in \cite{Niehoff:2012wu} to describe multiple superthreads with independent and arbitrary profiles. Upon smearing the multi superthread solutions give two-dimensional supersheet solutions that are in general characterized by profile functions of two variables. These solutions where found to be free of Dirac-strings without imposing any further constraints, in contrast to five-dimensional solutions where one writes the solution in terms of arbitrary electric and magnetic charges and then has to solve algebraic bubble equations that relate the charges and the positions of the objects. For a specific choice of profile functions the smearing integrals were performed to find a single supersheet solution with trivial reduction to five dimensions \cite{Bena:2004wt, Bena:2004wv, Bena:2005ni}. 

Here we generalize the results of \cite{Niehoff:2012wu} by considering corrugated multi-supersheet solutions that after smearing depend non-trivially in the compactification direction and are thus genuinely six-dimensional. Specifically we consider coaxial multi-supersheet geometries such that the shape of each supersheet is fixed but its scale varies within the compactification direction. They represent a new class of six-dimensional solutions that are by construction free of Dirac-strings. In five dimensions the cancellation of Dirac strings is given by integrability conditions which for single objects appear as a radius relation that gives the position of the object in terms of its charges, while for multiple objects they express the interactions between them in terms of magnetic fluxes. Thus it is interesting to explore how these interactions get embedded in the solution in the six-dimensional geometry and allow us in the future to construct microstate geometries in six dimensions. Analyzing the structure of these objects will also help us understand more about the superstratum which is also a genuine six-dimensional geometry. A novel feature of these new geometries is that, because of their fluctuation in the sixth dimension, under certain conditions they can touch or even intersect through each other.

In section 2 we briefly summarize the main results of \cite{Niehoff:2012wu} and extend the formalism to describe multi-supersheets. In section 3 for a specific choice of profile functions we calculate the smearing integrals to construct a corrugated supersheet, with an arbitrary periodic function describing the fluctuations in the compact direction. We further examine the asymptotic charges of the solution and the restrictions put on this arbitrary function by the absence of closed timelike curves. In section 4 we generalize the results of section 3 to construct multiple corrugated supersheets with independent and arbitrary oscillation profiles along the compact direction. We derive the local conditions that need to be satisfied for two supersheets to touch or intersect and by a specific choice of examples we provide a numerical analysis of the global conditions as well. In section 5, based on results of section 3, we give general arguments about the structure of six-dimensional solutions and a possible perturbative approach towards constructing black geometries as well as the superstratum. Finally we display our conclusions in section 6.

\section{Superthreads and Supersheets}\label{threadssheets}
In this section we describe superthreads and supersheets by reviewing the main results of \cite{Niehoff:2012wu} and providing some simple generalizations. The six-dimensional metric is
\bq \label{metric}
ds^2 = 2H^{-1}(dv+\beta)\left(du+\omg+\frac{\mathcal{F}}{2}(dv+\beta)\right) - Hds_4^2 .
\eq
The four-dimensional base metric satisfies some special conditions \cite{Bena:2011dd,Gutowski:2003rg} that will not be relevant here since we are simply going to take $ds_4^2$ to be the flat metric on $\IR^4$. The coordinate $v$ has period $L$ and the metric functions in general depend on both $v$ and the $\IR^4$ coordinates $\vec{x}$. We want to describe $1/8$-BPS geometries that have D1-D5-P charges. For simplicity we choose $\beta=0$, which means that there will be no Kaluza-Klein monopoles in the solution. In addition to that, because $\beta=0$, although the inhomogeneous terms in the BPS equations \cite{Niehoff:2012wu} contain $v$ derivatives, the differential operators in the left hand side contain no $v$ derivatives. The latter means that because of linearity, the solutions we find, although $v$-dependent, can be thought to be assembled by constant $v$ sections of the full solution.

\subsection{Multi-Superthreads}\label{multithreadssection}

\begin{figure}[!ht]
\begin{center}
\includegraphics[width=8.0cm]{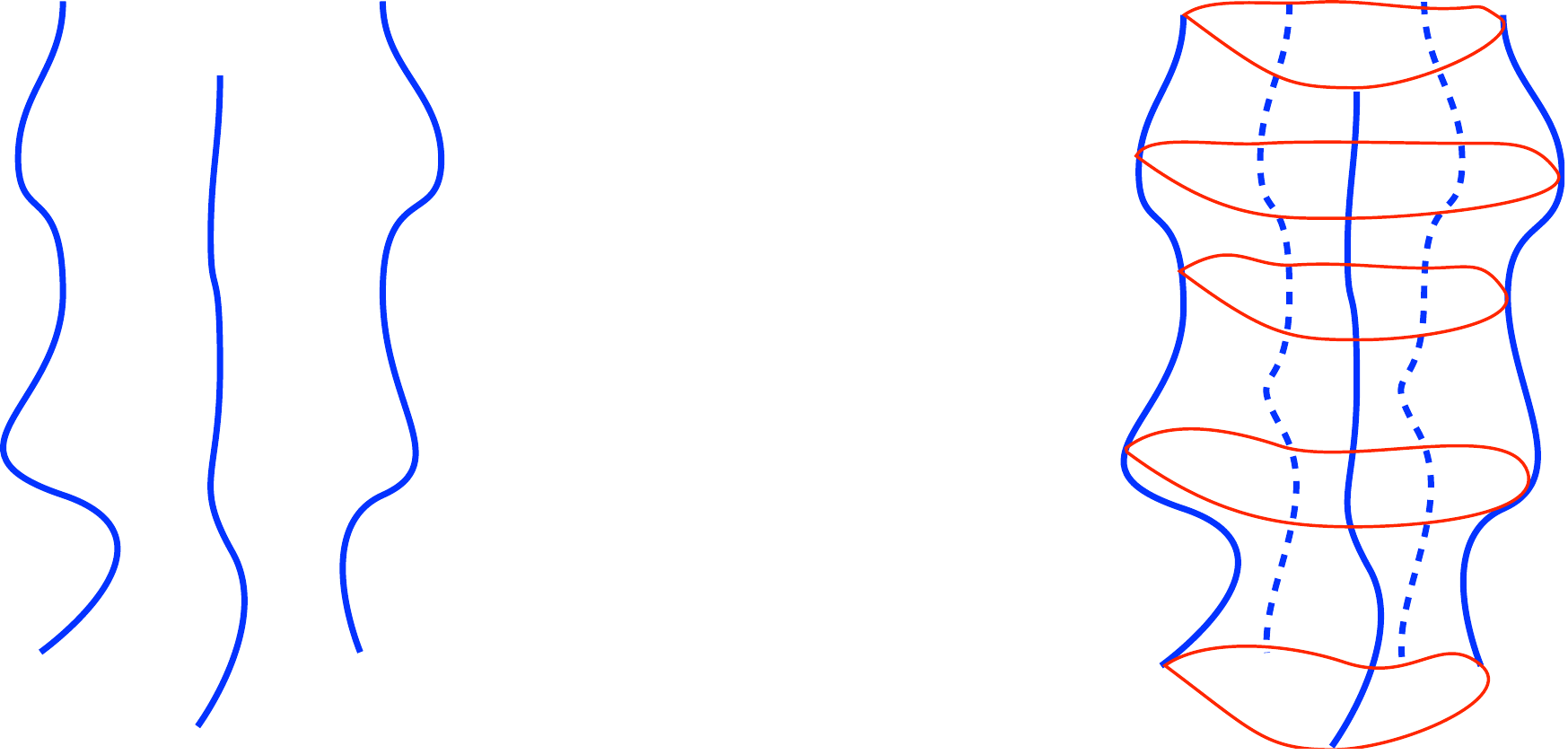}
\caption{ \small \it     Multi-superthread solution with independent profile functions $\vec{F}^{(p)}(v)$ for each thread.  After smearing the supersheet is described by generic functions of two variables $\vec{F}(\sigma,v)$.}
\label{fig0}
\end{center}
\end{figure}

The solutions consist of supersymmetric one-dimensional threads (hence superthreads) that run across the $v$ coordinate. The shape of the superthreads is given by profile functions $\vec{F}(v)$. In \cite{Niehoff:2012wu} the BPS equations were solved to describe multi-superthread solutions with independent and arbitrary shapes (fig.\ref{fig0}). The solution contains non-trivial shape-shape interaction terms which disappear when all the superthreads are parallel to each other.\\
Defining $H=\sqrt{Z_1Z_2}$ with $Z_1$,$Z_2$ harmonic functions that encode the $D1$, $D5$ charges respectively and denoting by $\vec{F}^{(p)}(v)$ the profile function of the p$^{th}$ thread, the multi-superthread solution has constituent parts
\bq \label{multithreads}
\begin{split}
& Z_m=  1 ~+~ \sum_{p=1}^n{Q_{m\, p}\over  |\vec x-\vec F^{(p)}(v) |^2} ,\\
& \mathcal{F}  ~=~  -4~-~ 4  \sum_{p=1}^n \,  \frac{Q_{3\, p}}{R_p^2}   ~-~  \frac{1}{2} \, \sum_{p, q=1  }^n    \, \frac{(Q_{1p} Q_{2q} + Q_{2p} Q_{1q})}{R_p^2 R_q^2} \, \Big( {\dot {\vec {F}}}{}^{(p)} \cdot {\dot {\vec {F}}}{}^{(q)}   \Big) +   \\
& + ~  \sum_{p, q=1 \atop p \ne q }^n  (Q_{1p} Q_{2q} + Q_{2p} Q_{1q})  \, \frac{1}{R_p^2 R_q^2} \,  \frac{  \dot{F}^{(p)}_i \, \dot{F}^{(q)}_j \,  \cA^{(p,q)}_{ij} }{\big | \vec F^{(p)}  - \vec F^{(q)} \big | ^2}  ,\\
& \omg=\omg_0 + \omg_1 + \omg_2 ,\\
&  \omega_0  ~=~    \sum_{m=1}^2 \,\sum_{p=1}^n{Q_{m\, p}\, \dot{F}^{(p)}_i dx^i \over |\vec x-\vec F^{(p)}(v) |^2}   ,\nonumber \\
& \omega_1  ~=~   \frac{1}{2}  \,\sum_{p, q=1}^n \, (Q_{1\, p}\,Q_{2\, q} +Q_{2\, p}\,Q_{1\, q})  \, { \dot{F}^{(p)}_i dx^i \over R_p^2  \,R_q^2} ,\\
& \omega_2 ~=~   \frac{1}{4}  \,\sum_{p, q=1 \atop p \ne q}^n  (Q_{1p} Q_{2q} + Q_{2p} Q_{1q}) \,  \frac{\big( \dot{F}^{(p)}_i - \dot{F}^{(q)}_i \big)  }{\big | \vec F^{(p)}  - \vec F^{(q)} \big | ^2} \, \bigg\{ \bigg( \frac{1}{R_p^2} - \frac{1}{R_q^2} \bigg)  \, dx^i
~-~ \frac{2}{R_p^2 R_q^2} \, \cA^{(p,q)}_{ij}  \, dx^j  \bigg\} ,
\end{split}
\eq
where in the above we used the notation $\partial_v \Phi \equiv \dot{\Phi}$. The charges $Q_{1p}$, $Q_{2p}$ and $Q_{3p}$ represent the D1, D5, P charges of the p$^{th}$ superthread respectively. Also we required that $Z_i\rightarrow 1$ at infinity so that the metric is asymptotically Minkowskian and we define
\bq \label{threaddefn}
\begin{split}
& \vec{R}_p \equiv \vec{R}^{(p)}=\vec{x}-\vec{F}^{(p)} ,\\
& \mathcal{A}^{(p,q)}_{ij}    ~\equiv~  R^{(p)}_i R^{(q)}_j   - R^{(p)}_j R^{(q)}_i  ~-~  \varepsilon^{ijk\ell} R^{(p)}_k R^{(q)}_\ell ,
\end{split}
\eq
with $\varepsilon^{1234}=1$. The anti-self-dual twoform area element $\mathcal{A}^{(p,q)}_{ij}$ encodes the non-trivial interactions between non-parallel superthreads.

\subsection{Multi-Supersheets}\label{multisheets}
In the continuum limit the summations are promoted to integrals and the indices $p$, $q$ become continuous variables $\sigma_1$, $\sigma_2$. The objects that occur after smearing are two-dimensional geometries which are called supersheets (fig.\ref{fig0}). The continuum limit for a single supersheet was considered in \cite{Niehoff:2012wu}. Here we give the immediate generalization for multiple supersheets with arbitrary and independent two-dimensional profiles. Thus for a multi-supersheet solution the profile functions $\vec{F}_p(v)$ become functions of two variables $\vec{F}_I(\sigma^{(I)},v)$ and the discrete charges $Q_{mp}$ are being replaced by density functions $\rho^{(I)}(\sigma^{(I)})$. In the above we used capital latin indices to separate between different supersheets of the solution. Then we have
\bq \label{multisheetint}
\begin{split}
& Z_m=  1 ~+~\sum_I \int_0^{2\,\pi} \, {\rho^{(I)}_m(\sigma^{(I)}) \, d \sigma^{(I)} \over  |\vec x-\vec F_I (\sigma^{(I)}, v) |^2} ,\\
& \mathcal{F}  ~=~  - 4~-~\sum_I 4\,\int_0^{2\,\pi}    \frac{ \rho^{(I)}_3(\sigma^{(I)}) }{R^{(I)} (\sigma^{(I)})^2}  \, d \sigma^{(I)} \\
& -~\sum_{I,J} \int_0^{2\,\pi}  \int_0^{2\,\pi} \,  (\rho^{(I)}_1(\sigma^{(I)}_1) \rho^{(J)}_2(\sigma^{(J)}_2)+\rho^{(I)}_2(\sigma^{(I)}_1) \rho^{(J)}_1(\sigma^{(J)}_2))  \   \frac{1}{R^{(I)} (\sigma^{(I)}_1)^2  R^{(J)} (\sigma^{(J)}_2)^2 } \\
&  \Bigg[\,  \frac{1}{2} \, \big( \partial_v  \vec F_I  (\sigma^{(I)}_1,v)\big)   \cdot \big(\partial_v  \vec F_J  (\sigma^{(J)}_2,v)  \big) ~-~    \frac{  \partial_v F_{I,i}(\sigma^{(I)}_1,v) \,  \partial_v F_{J,j} (\sigma^{(J)}_2,v) \ \,  \mathcal{A}^{(IJ)}_{ij}  (\sigma^{(I)}_1, \sigma^{(J)}_2) }{\big |  \vec  F_I (\sigma^{(I)}_1,v)   - \vec F_J (\sigma^{(J)}_2,v) \big |^2}  \, \Bigg]  \, d \sigma^{(I)}_1 d \sigma^{(J)}_2  ,\\
&  \omega_0  ~=~   \sum_I  \sum_{m=1}^2 \, \int_0^{2\,\pi} \, {\rho^{(I)}_m(\sigma^{(I)}) \, \partial_v  \vec  F_I(\sigma^{(I)},v) \cdot  d\vec x  \over |\vec x-\vec F_I (\sigma^{(I)}, v)  |^2}\, d \sigma^{(I)}  ,\\
&  \omega_1  ~=~    \frac{1}{2} \sum_{I,J} \,\int_0^{2\,\pi}  \int_0^{2\,\pi} \, {(\rho^{(I)}_1(\sigma^{(I)}_1) \rho^{(J)}_2(\sigma^{(J)}_2)+\rho^{(I)}_2(\sigma^{(I)}_1) \rho^{(J)}_1(\sigma^{(J)}_2))\partial_v  \vec  F_I(\sigma^{(I)}_1,v) \cdot  d\vec x \over R^{(I)} (\sigma^{(I)}_1, v, \vec x)^2  \, R^{(J)} (\sigma^{(J)}_2, v, \vec x)^2} \, d \sigma^{(I)}_1 d \sigma^{(J)}_2 ,\\
& \omega_2 ~=~     \frac{1}{4} \sum_{I,J}  \,\int_0^{2\,\pi}  \int_0^{2\,\pi}  (\rho^{(I)}_1(\sigma^{(I)}_1) \rho^{(J)}_2(\sigma^{(J)}_2)+\rho^{(I)}_2(\sigma^{(I)}_1) \rho^{(J)}_1(\sigma^{(J)}_2))  \,   \frac{\big( \partial_v  F_{I,i} (\sigma^{(I)}_1,v)  -   \partial_v  F_{J,i} (\sigma^{(J)}_2,v) \big)  } {\big |  \vec  F_I (\sigma^{(I)}_1,v)   - \vec F_J (\sigma^{(J)}_2,v) \big | ^2}  \\
&  \bigg\{ \bigg( \frac{1}{R^{(I)} (\sigma^{(I)}_1)^2} - \frac{1}{R^{(J)} (\sigma^{(J)}_2)^2} \bigg)  \, dx^i ~-~ \frac{2}{R^{(I)} (\sigma^{(I)}_1)^2  R^{(J)} (\sigma^{(J)}_2)^2 } \, \mathcal{A}^{IJ}_{ij}  (\sigma^{(I)}_1, \sigma^{(J)}_2)  \, dx^j  \bigg\} \, d \sigma^{(I)}_1 d \sigma^{(J)}_2 ,
\end{split}
\eq
where we define
\bq \label{sheetdefn}
\begin{split}
& \vec R^{(I)} (\sigma^{(I)})   ~\equiv~  \vec x ~-~ \vec F_I (\sigma^{(I)}, v) ,\\
& \mathcal{A}^{IJ}_{ij}  (\sigma^{(I)}_1, \sigma^{(J)}_2)    ~\equiv~  R^{(I)}_i (\sigma^{(I)}_1)   R^{(J)}_j (\sigma^{(J)}_2)  -  R^{(I)}_j (\sigma^{(I)}_1)   R^{(J)}_i  (\sigma^{(J)}_2) 
   ~-~  \varepsilon^{ijk\ell} R^{(I)}_k (\sigma^{(I)}_1)  R^{(J)}_\ell   (\sigma^{(J)}_2) .
\end{split}
\eq
The capital latin indices in the integration variables $\si_i^{(I)}$ of (\ref{multisheetint}), (\ref{sheetdefn}) are dummy indices and can be removed. For the single integrals this is a trivial observation, while for the double integrals it is based on the symmetry of the integrand in exchanging $I\leftrightarrow J$. The structure of the equations (\ref{multisheetint}) indicates that a mult-supersheet solution will consist of two parts. First, there will be a summation over individual supersheets of different radii and profiles coming from single summation terms and from double summations when $I=J$. Secondly there will be interaction terms between the different supersheets coming from double summations when $I\neq J$. Furthermore, because superthreads interact in pairs so will the resulting supersheets. Thus the solution naturally decomposes as
\bq \label{multisheetdecomp}
\begin{split}
& Z_i=1+\sum H^{(I)}_i ,\\
& \omg=\sum_I \omg^{(I)} + \sum_{I\neq J} \omg^{(I,J)} ,\\
& \mathcal{F}=-4+\sum_I\mathcal{F}^{(I)} + \sum_{I\neq J}\mathcal{F}^{(I,J)} .
\end{split}
\eq
Because the integrands of $\omg_2$ and $\mathcal{F}$ are symmetric in exchanging $I\leftrightarrow J$, we have $\omg_2^{(I,J)}=\omg_2^{(J,I)}$ and $\mathcal{F}^{(I,J)}=\mathcal{F}^{(J,I)}$. So we can further reduce (\ref{multisheetdecomp}) to
\bq \label{multisheetdecomp2}
\begin{split}
& Z_i=1+\sum H^{(I)}_i ,\\
& \omg= \sum_I \omg^{(I)} + \sum_{I< J} \left( \omg_1^{(I,J)}+\omg_1^{(J,I)}+2\omg_2^{(I,J)} \right) ,\\
& \mathcal{F}=-4+\sum_I\mathcal{F}^{(I)} + 2\sum_{I< J}\mathcal{F}^{(I,J)} .
\end{split}
\eq
%

\section{A Corrugated Supersheet} \label{singlebreather}
In \cite{Niehoff:2012wu} by considering superthreads of helical profile the smearing integrals for a single supersheet were explicitly calculated. Although the profile functions depended on both $v$ and $\sigma$ the resulting supersheet was independent of $v$ and it matched a special class of already known five-dimensional solutions \cite{Bena:2004wt}, \cite{Bena:2004wv}, \cite{Bena:2005ni}. Here we want to extend these results by considering a slightly more general choice of profile functions so that the resulting supersheet can also fluctuate in the coordinate $v$. Thus, we choose
\bq \label{profile}
\vec{F}(v,\sigma)=(A(\lm v) \cos(\ka v+\si)\, , \, A(\lm v) \sin(\ka v+\si)\, , \, 0 \, , \, 0) .
\eq
As in  \cite{Niehoff:2012wu} we once again choose helical profile functions, but now the radius $A$, instead of being constant, is an arbitrary function of $v$. The resulting supersheet will have a  circular profile in $\IR^4$ for a specific value of $v$, with varying circle radius as we move along the $v$ direction. The constants $\lm$ and $\ka$ are of the form $2\pi n/L$, where $n$ is an integer, and can in principle be independent. Although our results are for a circular profile we believe that they can be generalized to any closed,  non-intersecting curve by constructing the appropriate Green's functions.\\
For the $\IR^4$ base space with take double polar coordinates
\bq \label{polarcoord}
x_1=\eta \cos\psi \, , \, x_2=\eta \sin\psi \, , \, x_3=\zeta \cos\phi \, , \,x_4=\zeta \sin\phi ,
\eq
from which we can go to spherical coordinates by substituting
\bq \label{sphericalcoord}
\eta=r\sin\theta \, , \, \zeta=r\cos\theta .
\eq
%

\subsection{The solution} \label{singlebreathersolution}
To describe a single supersheet we need to calculate the integrals (\ref{multisheetint}) with no summations and the latin indices removed. We also consider a constant charge distribution
\bq \label{singlechargedensity}
\rho_m(\sigma)=\frac{Q_m}{2\pi} .
\eq
 The integrals are easily calculated by going to the complex plane and summing the residues of the poles that are within the unit circle. The double integrals transform to a double complex integral and the residues of the first complex integration act as integrand functions for the second complex integral. Then for the functions describing the solution we find
\bq \label{singlez}
Z_1=1+\frac{Q_1}{\Sigma} \, , \, Z_2=1+\frac{Q_2}{\Sigma} ,
\eq
where
\bq \label{sigma}
\Sigma=\sqrt{(A^2+\eta^2+\zeta^2)^2-4A^2\eta^2}= \sqrt{(A^2+r^2)^2-4A^2r^2\sin^2\theta} ,
\eq
is the position of the supersheet in $\IR^4$ and is now $v$ dependent via the function $A(\lm v)$.\\
Similarly we have
\bq \label{singleF}
\mathcal{F}=-4- \frac{4Q_3A^2-Q_1Q_2\left(\lm^2\ac{A}^2+\ka^2A^2\right)}{A^2\Sigma}-Q_1Q_2\frac{\left(\lm^2\ac{A}^2+\ka^2A^2\right)\left(\eta^2+\zeta^2\right)}{A^2\Sigma^2}.
\eq
For the angular momentum one obtains
\bq \label{singleomg}
\begin{split}
& \omg=\frac{(Q_1+Q_2)}{2\eta A}\left( -1 + \frac{A^2+\eta^2 + \zeta^2}{\Sigma} \right)\left( \ka A \eta d\psi + \lm\ac{A} d\eta \right) +  \\
& +Q_1Q_2 \frac{\eta}{A\Sigma^2} \left( \eta \ka A d\psi + \lm \ac{A} d\eta \right) + Q_1Q_2 \frac{\zeta}{A\Sigma^2} \left( \zeta \ka A d\phi + \lm \ac{A} d\zeta \right),
\end{split}
\eq
where
\bq \label{Aderiv}
\partial_vA=\dot{A}=\ac{A}\lm.
\eq
It is interesting to observe that the functions (\ref{singleF}) and (\ref{singleomg}) 
, because of linearity of the BPS equations, can be considered as the ``superposition of modes" occurring from two distinct cases of supersheets: one made from superthreads of helical profile with constant radius ($\lm=0$) and another made of straight superthreads with a corrugated profile ($\ka=0$). Thus corrugated and helical modes of the solution are independent, originating from the fact that the profile function factorizes and each mode is sourced by a different factor of (\ref{profile}).

 The $\IR^4$ coordinates on which the angular momentum one form $\omg$ has legs depend on $\dot{F}_idx^i$, (\ref{multisheetint}). Thus the helical mode generates the components $d\psi$ and $d\phi$. The corrugated mode generates new components $d\eta$ and $d\zeta$ along radial directions, expressing that the $\IR^4$ circular profile changes radius as we move along $v$.

\subsection{Regularity and Asymptotic Charges} \label{regularity}

Before examining our solution for closed timelike curves there are some restrictions to be placed on the function $A$. In general there is the possibility of antipodal points of the $\IR^4$ circular profile to intersect over each other at specific values of $v=v_0$. For our choice of profile functions this happens exactly when $A(\lm v_0)=0$. Then the functions $\mathcal{F}$ and $\omg$ diverge. Thus we need
\bq \label{Aconstraint}
A(\lm v)\neq 0 \quad \forall v \in [0,L] .
\eq
For example for the simple choice $A=b+a\cos(\lm v)$ we need $b>a$. However, since $A$ is a periodic function of $v$ there will be points $v=v_1$ such that $\dot{A}(\lm v_1)=0$. For these values of $v$ the solution and its physical analysis exactly matches that of \cite{Niehoff:2012wu} where $A$ is constant. That might lead to an interesting perturbative approach in exploring the superstratum by expanding around the points $\dot{A}(\lm v_1)=0$ where the superstratum should match the $v$-independent supertube solution. Thus the superstratum can possibly be realized as additional perturbation modes in the supertube solution around these points. We further comment on this idea in section \ref{KKM}. A different perturbative approach to the superstratum \footnote{The superstratum discussed in \cite{Giusto:2012jx} would be a generalization of the three electric one, magnetic dipole charge supertube presented in \cite{Vasilakis:2012zg}.}, which differs from what we discuss here has recently appeared in \cite{Giusto:2012jx}.\\

We read the asymptotic charges of the solution from the expansion of $Z_1$, $Z_2$, $\mathcal{F}$ and $\omg$ for $r\rightarrow \infty$. The asymptotic electric charges are
\bq \label{asympelectric}
Q_{1,\infty}=Q_1 \, , \, Q_{2,\infty}=Q_2 \, , \, Q_{3,\infty}=Q_3
\eq
and from the expansion of $\omg$ we get
\bq \label{omgexp}
\omg \sim \frac{1}{r^2}\left(\left(\ka A^2(Q_1+Q_2)+\ka Q_1Q_2\right)\sin^2\theta d\psi + \ka Q_1Q_2\cos^2\theta d\phi + \frac{1}{2}(Q_1+Q_2)\lm A \dot{A}\sin(2\theta)d\theta \right).
\eq
Thus, in contrast to the case that the supersheet is $v$-independent there is an additional term proportional to $\dot{A}$ along the $\theta$ direction.

To examine the solution for closed timelike curves it is useful to rewrite the metric (\ref{metric}) by completing the squares
\bq \label{metric2}
ds^2=H^{-1}\mathcal{F}\left(dv+\beta + \mathcal{F}^{-1}(du+\omg)\right)^2 - H^{-1}\mathcal{F}^{-1}\left(du+\omg\right)^2 - Hds_4^2 .
\eq
Taking a slice of $u=$constant the absence of closed timelike curves requires that the following conditions hold globally
\bq \label{noCTC1}
\mathcal{F}\le 0 ,
\eq
\bq \label{noCTC2}
ds_4^2+\frac{\omg^2}{H^2\mathcal{F}} \ge 0 .
\eq
From (\ref{noCTC1}) for $r\rightarrow 0$ we get
\bq \label{Freg}
Q_3 \ge \frac{Q_1Q_2\left(\lm^2\ac{A}^2+\ka^2A^2\right)}{4A^2} .
\eq
The charge $Q_3$ should be big enough so that is greater than the right hand side of (\ref{Freg}) for every value of $v$. Since $Q_3$ enters $\mathcal{F}$ as a harmonic term, the charge $Q_3$ can be an arbitrary function of $v$ without affecting the solution. Consequently by defining
\bq \label{Q3hat}
Q_3(v)=\widehat Q_3 \frac{\left(\lm^2\ac{A}^2+\ka^2A^2\right)}{\ka^2A^2} ,
\eq
we get a result similar to the five dimensional case examined in \cite{Niehoff:2012wu}
\bq \label{Freg2}
\widehat Q_3\ge \frac{1}{4}\ka^2Q_1Q_2 ,
\eq
where $\widehat Q_3$ is an effective $v$-independent charge. From (\ref{noCTC2}) by taking the near supersheet limit $\Sigma\rightarrow 0$ we obtain from the leading order term
\bq \label{CTCconstraint}
\frac{1}{\ka^2 A^2 + \lm^2 \ac{A}^2}\left( A \ka dr -\ac{A}\lm r \sin^2\theta d\psi - \ac{A}\lm r \cos^2\theta d\phi \right)^2 + r^2\sin^2\theta \cos^2\theta \left(d\psi + d\phi \right)^2 + rd\theta^2 ,
\eq
%
%
%
which is always positive. Suppose we have $\dot{A}(\lm v_1)=0$ at some point $v_1$, then we can make the leading order term vanish by choosing, consistently with the $\Sigma \ra 0$ limit, $r\ra A$, $\theta\ra \pi/2$. By considering the next to leading order term in the expansion we get the constraint given in \cite{Niehoff:2012wu} for a non-corrugated supersheet, which is
\bq \label{CTCconstraint2}
Q_1Q_2\left(\widetilde Q_3(v_1) - \frac{1}{4}\ka^2A(\lm v_1)(Q_1+Q_2) \right) \ge 0 .
\eq
Thus
\bq \label{chargemomentum}
\widetilde Q_3(v_1) \ge  \frac{1}{4}\ka^2A(\lm v_1)(Q_1+Q_2) ,
\eq
where we defined
\bq \label{Q3tilde}
\widetilde Q_3(v)=Q_3 - \frac{Q_1Q_2\left(\lm^2\ac{A}^2+\ka^2A^2\right)}{4A^2} .
\eq
Also in the limit $\Sigma \ra 0$ the metric function blows up and thus the corrugated supersheet geometry is singular. This is consistent with the five-dimensional non-corrugated supersheet picture and the comment in the beginning of section \ref{threadssheets} that because $\beta=0$ these geometries can be thought as a collection of slices of constant $v$. Thus for every value of $v$ we have a singularity with radial profile in $\IR^4$ and the collection of all the different $v$-slices creates a singular six-dimensional geometry.

\section{Corrugated Multi-Supersheets} \label{multibreather}
Here we calculate the smearing integrals (\ref{multisheetint}) for corrugated multi-supersheets. Generalizing (\ref{profile}) we separate the superthreads at different sets $A_I$ with profile functions at each set
\bq \label{multiprofile}
 \vec{F}_{I}(v,\sigma^{(I)})=(A_I(\lm_I v) \cos(\ka_I v+\si^{(I)})\, , \, A_I(\lm_I v) \sin(\ka_I v+\si^{(I)})\, , \, 0 \, , \, 0) ,
\eq
with $A_I<A_J$ for $I<J \quad \forall v\in[0,L]$.\\
To each set of threads we assign a constant charge distribution given by
\bq \label{multicharge}
\rho_i^{(I)}(\si^{(I)})=\frac{Q_i^{(I)}}{2\pi} .
\eq
%

\subsection{The Solution} \label{multibreathersolution}

 These supersheets have concentric circular profiles in $\IR^4$ with radii $A_I(\lm_I v)$ which fluctuate as we move along the $v$ coordinate. In general they are also non-parallel as each one has its own function $A_I(\lm_Iv)$ that oscillates along $v$. By using (\ref{multisheetdecomp2}) we write the solution as a combination of individual supersheets and interaction terms. For the individual supersheets terms we have
 \bq \label{multisingle}
 \begin{split}
& H_m^{(I)}=\frac{Q_m^{(I)}}{\Sigma_I} ,\\
& \mathcal{F}^{(I)}=- \frac{4Q^{(I)}_3A_I^2-Q^{(I)}_1Q^{(I)}_2\left(\lm_I^2\ac{A_I}^2+\ka_I^2A_I^2\right)}{A_I^2\Sigma_I}-Q^{(I)}_1Q^{(I)}_2\frac{\left(\lm_I^2\ac{A_I}^2+\ka_I^2A_I^2\right)\left(\eta^2+\zeta^2\right)}{A_I^2\Sigma_I^2} ,\\
& \omg^{(I)}=\frac{(Q^{(I)}_1+Q^{(I)}_2)}{2\eta A_I}\left( -1 + \frac{A_I^2+\eta^2 + \zeta^2}{\Sigma_I} \right)\left( \ka_I A_I \eta d\psi + \lm_I\ac{A}_I d\eta \right) +  \\
& +Q^{(I)}_1Q^{(I)}_2 \frac{\eta}{A_I\Sigma_I^2} \left( \eta \ka_I A_I d\psi + \lm_I \ac{A}_I d\eta \right) + Q^{(I)}_1Q^{(I)}_2 \frac{\zeta}{A_I\Sigma_I^2} \left( \zeta \ka_I A_I d\phi + \lm_I \ac{A}_I d\zeta \right) \end{split}
 \eq
 and for the interaction terms

\bq \label{multiinteraction}
\begin{split}
& \mathcal{F}^{(I,J)}= -\frac{4Q_3^{(I,J)}}{\Sigma_I}- \frac{Q_1^{(I)}Q_2^{(J)}+Q_2^{(I)}Q_1^{(J)}}{2} \left( \lm_I\lm_J \ac{A_I}\ac{A_J} + \ka_I\ka_J A_IA_J \right) \left( - \frac{1}{A_IA_J\Sigma_I} + \frac{\eta^2+\zeta^2}{A_IA_J\Sigma_I\Sigma_J} \right) ,\\
& \omg_1^{(I,J)}=\frac{Q_1^{(I)}Q_2^{(J)} + Q_2^{(I)}Q_1^{(J)}}{4\eta A_J \Sigma_I \Sigma_J}\left(A_J^2+\eta^2+\zeta^2 -\Sigma_J\right)\left(\ka_JA_J\eta d\psi + \lm_J \ac{A}_Jd\eta \right) ,\\
& \omg_{2}^{(I,J)}=\left(Q_1^{(I)}Q_2^{(J)} + Q_2^{(I)}Q_1^{(J)}\right) \Bigg( \left(- G_1^{(I,J)}\ka_I + G_2^{(I,J)}\ka_J\right)d\psi + \\
& + \left(-\frac{\ac{A}_I}{A_I\eta}G_1^{(I,J)}\lm_I + \frac{\ac{A}_J}{A_J\eta}G_2^{(I,J)}\lm_J\right) d\eta + \\
& + \left( L_1^{(I,J)}A_IA_J(\ka_I+\ka_J) + L_2^{(I,J)}(\ka_IA_I^2+\ka_JA_J^2)\right)d\phi +\\
& + \frac{1}{\zeta}\left( L_1^{(I,J)}(\lm_IA_J\ac{A}_I+\lm_JA_I\ac{A}_J) + L_2^{(I,J)}(\lm_IA_I\ac{A}_I+\lm_JA_J\ac{A}_J) \right)d\zeta \Bigg) ,
\end{split}
\eq
where as in (\ref{multisheetdecomp2}) we require $I<J$ and we also define
\bq \label{multisigma}
\begin{split}
& \Sigma_I=\sqrt{(A_I^2+\eta^2+\zeta^2)^2-4A_I^2\eta^2}= \sqrt{(A_I^2+r^2)^2-4A_I^2r^2\sin^2\theta}
\end{split}
\eq
and we have introduced the functions

\bq \label{multidefn}
\begin{split}
& G_1^{(I,J)}= \frac{(A_I^2+\Sigma_J + \zeta^2)(A_I^2-\Sigma_I+\zeta^2)+\eta^2(2\zeta^2-2A_I^2+\Sigma_J-\Sigma_I) +\eta^4}{8(A_I^2-A_J^2)\Sigma_I\Sigma_J} ,\\
& G_2^{(I,J)}=  \frac{A_J^4+2A_I^2\Sigma_J-(\Sigma_I-\eta^2-\zeta^2)(\Sigma_J+\eta^2+\zeta^2) - A_J^2(\Sigma_I + \Sigma_J +2\eta^2-2\zeta^2) }{8(A_I^2-A_J^2)\Sigma_I\Sigma_J} ,\\
& L_1^{(I,J)} = \frac{A_IA_J\zeta^2\left( A_I^2(\Sigma_J-2\eta^2) + (\Sigma_I+\Sigma_J)(\eta^2+\zeta^2)+A_J^2(\Sigma_I+2\eta^2) \right)}{(A_I^2-A_J^2)\Sigma_I\Sigma_J(A_J^2-A_I^2+\Sigma_I+\Sigma_J)(A_I^2(\Sigma_J-\eta^2-\zeta^2)+A_J^2(\Sigma_I+\eta^2+\zeta^2))} ,\\
& L_2^{(I,J)} = \frac{\zeta^2\left( (A_I^2-A_J^2)\Sigma_I\Sigma_J - (A_I^2+A_J^2)(A_I^2+\eta^2+\zeta^2)\Sigma_J \right)}{2(A_I^2-A_J^2)\Sigma_I\Sigma_J(A_J^2-A_I^2+\Sigma_I+\Sigma_J)(A_I^2(\Sigma_J-\eta^2-\zeta^2)+A_J^2(\Sigma_I+\eta^2+\zeta^2))} - \\
& - \frac{\zeta^2\left((A_J^2+\eta^2+\zeta^2)\left( A_J^2(\Sigma_I+\eta^2+\zeta^2)+A_I^2(A_J^2+\Sigma_I-\eta^2-\zeta^2) - A_I^4 \right)  \right)}{2(A_I^2-A_J^2)\Sigma_I\Sigma_J(A_J^2-A_I^2+\Sigma_I+\Sigma_J)(A_I^2(\Sigma_J-\eta^2-\zeta^2)+A_J^2(\Sigma_I+\eta^2+\zeta^2))} .\\
\end{split}
\eq
Although the integrands of the interaction terms $\omg_2^{(I,J)}$ and $\mathcal{F}^{(I,J)}$ are symmetric under $I\leftrightarrow J$, the functions that occur after smearing are not. This is due to the fact that while calculating the integrals we had to make use of $A_I<A_J$ for $I<J$. Our solutions match the already studied five-dimensional solutions provided we set all $\lm_I=0$. In the five-dimensional case however one writes down the solution with freely adjustable parameters and positions and the interaction terms are symmetric under $I\leftrightarrow J$. Then one has to separately solve integrability conditions (also called bubble equations), coming from requiring the absence of Dirac strings, that relate the positions of the objects with their electric and dipole magnetic charges. Here the solution is by construction free of Dirac strings and that is reflected in the lack of symmetry under $I\leftrightarrow J$.\\

To this end it is useful to provide additional details on how the asymmetry in $I\leftrightarrow J$ arises from the multi-supersheet integrals (\ref{multisheetint}). To calculate any double integral expressing the interactions between the supersheets $I$ and $J$ we go to the complex plane by introducing complex variables $z=e^{i\si_1}$, $w=e^{i\si_2}$. We perform a double complex integration where the residues of the first integral become the integrand of the second. At each integral we pick the poles that are within the unit circle. The denominator of the integrand contains the factor $(A_Iz-A_Jw)(A_Jz-A_Iw)$, coming from the term $|\vec{F}_I(\si_1,v)-\vec{F}_J(\si_2,v)|^2$ in (\ref{multisheetint}). As a result when performing the first integral (either over $z$ or $w$) we need to know which of the two ratios $\frac{A_I}{A_J}$ or $\frac{A_J}{A_I}$ is less than one, to pick the pole within the unit circle, and thus the ordering of the supersheets matters.

\subsection{Regularity and Asymptotic Charges} \label{multiasymp}
Once again we have to verify that our solution is free of closed timelike curves (\ref{noCTC1}), (\ref{noCTC2}). For each of the terms of the solution representing an individual supesheet the analysis of section \ref{regularity} can be repeated in exactly the same manner and the  conditions (\ref{noCTC1}) and (\ref{noCTC2}) give exactly the conditions (\ref{Freg}) - (\ref{Q3tilde}) with the capital latin index $I$ included to specify the supersheet we are referring to. In the multi-supersheet case though one has to examine the interaction terms as well. Imposing
\bq\label{FinterCTC}
\mathcal{F}^{(I,J)} \le 0 ,
\eq
we get from (\ref{multiinteraction}) the additional constraint
\bq \label{Q3intercondition}
Q_3^{(I,J)} \ge \left(Q_1^{(I)}Q_2^{(J)} + Q_2^{(I)}Q_1^{(J)}\right)\frac{\lm_I\lm_J \ac{A_I}\ac{A_J} + \ka_I\ka_J A_IA_J }{8A_IA_J} .
\eq
Similarly to (\ref{Q3hat}) we can define an effective $v$-independent interaction charge $\widehat{Q}_3^{(I,J)}$ such that
\bq \label{Q3effinter}
Q_3^{(I,J)}(v)=\widehat{Q}_3^{(I,J)}\frac{\lm_I\lm_J \ac{A_I}\ac{A_J} + \ka_I\ka_J A_IA_J }{\ka_I\ka_JA_IA_J} .
\eq
Then (\ref{Q3intercondition}) reduces to the condition we would take for the five-dimensional case ($\lm_I=\lm_J=0$)
\bq\label{Q3effcondition}
\widehat{Q}_3^{(I,J)} \ge \left(Q_1^{(I)}Q_2^{(J)} + Q_2^{(I)}Q_1^{(J)}\right)\frac{\ka_I\ka_J  }{8} .
\eq
One should also verify that the global condition (\ref{noCTC2}) is satisfied by numerically examining the space between the different supersheets for specific choices of the functions $A_I(v)$. We partially perform this analysis later and find there are areas of the parameter space of solutions for which (\ref{noCTC2}) is satisfied. In general, because of the correspondence to the five-dimensional case, if the supersheets are well separated and the oscillations along $v$ are small we expect the solution to be regular.
\\

The asymptotic charges of the solution, in addition to being the sum of the individual supersheet pieces, will also get contributions from the interaction terms. The asymptotic electric charges are
\bq \label{multielectric}
Q_{1,\infty}=\sum_I Q_1^{(I)} \, , \, Q_{2,\infty}=\sum_I Q_2^{(I)} \, , \,  Q_{3,\infty}=\sum_I Q_3^{(I)} + 2\sum_{I<J}Q_3^{(I,J)} .
\eq
By expanding $\omg$ we get
\bq \label{multiomgexpansion}
\omg \sim \frac{1}{r^2}\left( \sum_I J^{(I)} + \sum_{I<J} J^{(I,J)}  \right) ,
\eq
where
\bq \label{multiomgexpansioncomps}
\begin{split}
& J^{(I)}=  \left( \ka_I A_I^2(Q^{(I)}_1+Q^{(I)}_2)+\ka_I Q^{(I)}_1Q^{(I)}_2\right) \sin^2\theta d\psi + \ka_I Q^{(I)}_1Q^{(I)}_2\cos^2\theta d\phi + \\
& +  \frac{1}{2}(Q^{(I)}_1+Q^{(I)}_2)\lm_I A_I \ac{A}_I \sin(2\theta)d\theta  ,\\
& J^{(I,J)}=\left(Q_1^{(I)}Q_2^{(J)}+Q_2^{(I)}Q_1^{(J)} \right)\ka_J\left(\sin^2\theta d\psi + \cos^2\theta d\phi \right) .
\end{split}
\eq
The contributions to the angular momentum that come from the interaction pieces neither depend on $\dot{A}$ nor do they have a $d\theta$ component. Thus the interaction terms in the asymptotic charges for corrugated multi-supersheets are the same with those for solutions with trivial reduction to five dimensions. It will be interesting to examine whether this holds when we include the third dipole charge i.e. $\beta \neq 0$. \\

Having fully described the solution there is an interesting coincidence limit where all the supersheets become identical
\bq \label{coincidence}
A_I\ra A \, , \, \lm_I\ra \lm \, , \, \ka_I \ra \ka \, , \, Q_i^{(I)}\ra Q_i \, , \, Q_3^{(I,J)} \ra Q_3 ,
\eq
for all the $N$ supersheets of the solution and for $i=1,2,3$. Then we obtain the single supersheet solution of section \ref{singlebreather} and its asymptotic charges with
\bq \label{coincidencecharges}
Q_1\ra NQ_1\, , \, Q_2\ra NQ_2 \, , \, Q_3\ra N^2Q_3 .
\eq
%


\subsection{Touching, Intersecting \& Regularity} \label{touching}
\begin{figure}[!ht]
\begin{center}
\includegraphics[width=12.0cm]{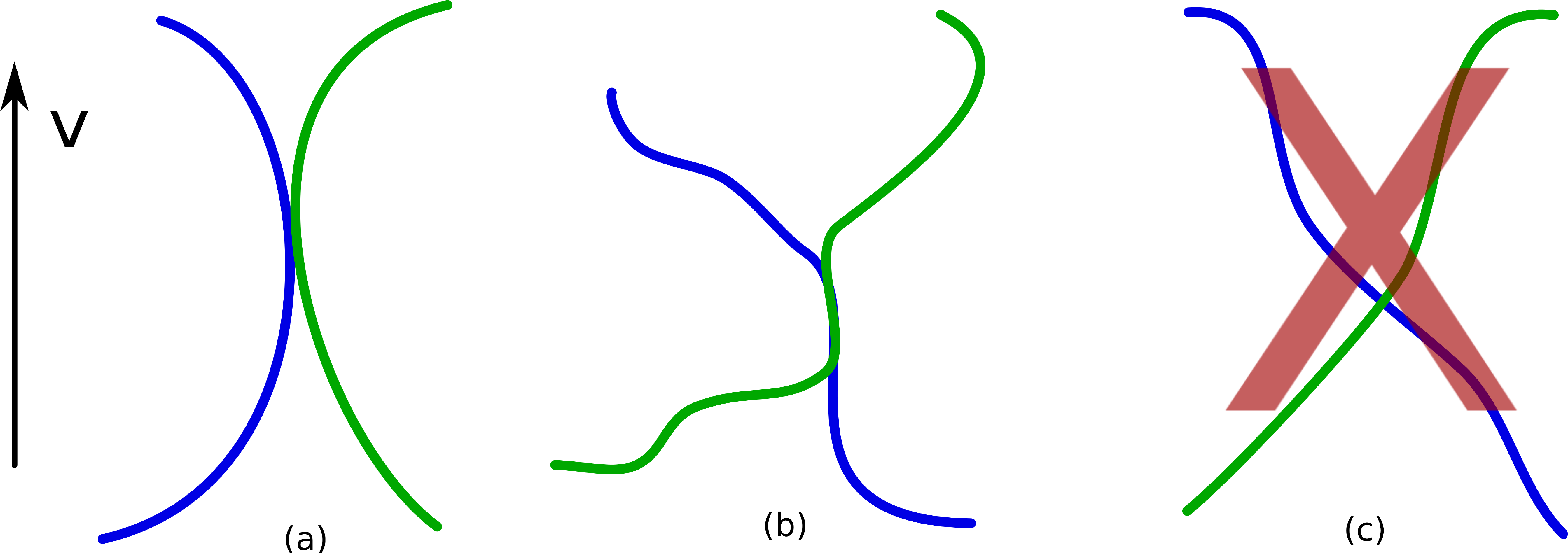}
\caption{ \small \it  Sections of different supersheets (blue \& green): (a) touching supersheets, (b) supersheets intersecting by touching tangentially, (c) supersheets cannot intersect without touching tangentially.}
 \label{fig1}
\end{center}
\end{figure}

Although so far we have strictly  imposed the condition $A_I<A_J\quad \forall v\in [0,L]$, the functions describing the solution are regular in the limit $A_I\ra A_J$. The individual supersheet terms $H^{(I)}$, $\mathcal{F}^{(I)}$, $\omg^{(I)}$ and the interaction terms $\mathcal{F}^{(I,J)}$, $\omg_1^{(I,J)}$ are trivially regular in this limit. For the terms appearing in $\omg_2^{(I,J)}$ the functions $G_1^{(I,J)}$ and $G_2^{(I,J)}$ are regular while $L_1^{(I,J)}$ and $L_2^{(I,J)}$ are singular. However one can observe from $\omg_2^{(I,J)}$ in (\ref{multiinteraction}) that in the limit $A_I\ra A_J$ the $d\phi$ and $d\zeta$ components factorize and the quantity $L_1^{(I,J)}+L_2^{(I,J)}$ is regular as well. Thus it seems that since supersheets can be realized as a collection of different $v$-slices we can have different supersheets touching or intersecting (fig.\ref{fig1}) through each other at specific values of $v=v_2$ such that $A_I(\lm_Iv_2)=A_J(\lm_Jv_2)$.  

For touching supersheets we should by definition have
\bq \label{touchingconditions1}
A_I(\lm_Iv_2)=A_J(\lm_Jv_2) \quad , \quad \dot{A}_I(\lm_Iv_2)=\dot{A}_J(\lm_Jv_2) .
\eq
Then continuity of the solution requires that taking the limit (\ref{touchingconditions1}) on the result (\ref{multiinteraction}), (\ref{multidefn}) should match the result of the smearing integral for $v=v_2$ where (\ref{touchingconditions1}) holds before smearing. Then one gets that the helical winding numbers of the two supersheets should match i.e. $\ka_I=\ka_J$. Overall the conditions for two supersheets touching at $v=v_2$ are
\bq \label{touchingconditions2}
\ka_I=\ka_J \, , \, A_I(\lm_Iv_2)=A_J(\lm_Jv_2) \, , \, \dot{A}_I(\lm_Iv_2)=\dot{A}_J(\lm_Jv_2) .
\eq

The somewhat notorious case of supersheets intersecting through each other such that $A_I(\lm_Iv_2)=A_J(\lm_Jv_2)$ needs some additional attention, as the ordering of the supersheets changes. As we mentioned at the end of section \ref{multibreathersolution} not only the functions $\mathcal{F}^{(I,J)}$ and $\omg_2^{(I<J)}$ describing the interactions between different supersheets are not symmetric under $I\leftrightarrow J$, but also the ordering of the supersheets matters in calculating the integrals. Reminding ourselves that our solution is a collection of different constant $v$ slices the previous issue could easily be resolved. One would have to calculate the smearing integrals of the interaction at two different areas of the coordinate $v$. For $v<v_2$ we have $A_I<A_J$ and the result would be the functions $\mathcal{F}^{(I,J)}$ and $\omg_2^{(I,J)}$ as given by (\ref{multiinteraction}), (\ref{multidefn}). For $v>v_2$ we would have  $A_I>A_J$ and thus
\bq \label{aftercrossing}
\begin{split}
&  \widetilde{\mathcal{F}}^{(I,J)} = \mathcal{F}^{(J,I)} ,\\
& \widetilde{\omg}_2^{(I,J)} = \omg_2^{(J,I)} .
\end{split}
\eq
Then we should at least require that the functions describing the supersheets are continuous at the intersection point $v=v_2$, which means we should demand
\bq \label{crossingcondition}
\begin{split}
& \lim_{v\to v_2}\left( \mathcal{F}^{(I,J)} - \widetilde{\mathcal{F}}^{(I,J)} \right) =0 ,\\
& \lim_{v\to v_2} \left( \omg_2^{(I,J)} - \widetilde{\omg}_2^{(I,J)} \right)=0 .
\end{split}
\eq
Using (\ref{aftercrossing}) we find for the difference of the functions
\bq \label{crossingdisaster}
\begin{split}
& \lim_{v\to v_2} \left(\mathcal{F}^{(I,J)} - \widetilde{\mathcal{F}}^{(I,J)} \right) = 4\frac{Q_3^{(J,I)}(v_2) - Q_3^{(I,J)}(v_2)}{\Sigma_I} ,\\
& \lim_{v\to v_2} \left( \omg_2^{(I,J)} - \widetilde{\omg}_2^{(I,J)} \right) =  \lim_{v\to v_2} \left(Q_1^{(I)}Q_2^{(J)} + Q_2^{(I)}Q_1^{(J)}\right) \cdot\\
& \cdot \Bigg( \left(- \left(G_1^{(I,J)}+G_2^{(J,I)}\right)\ka_I + \left(G_2^{(I,J)} + G_1^{(J,I)}\right)\ka_J\right)d\psi + \\
& + \left(-\frac{\ac{A}_I}{A_I\eta}\left(G_1^{(I,J)}+G_2^{(J,I)}\right)\lm_I + \frac{\ac{A}_J}{A_I\eta}\left(G_2^{(I,J)} + G_1^{(J,I)}\right)\lm_J\right) d\eta \Bigg) .
\end{split}
\eq
Thus taking the limit $v\ra v_2$, for (\ref{crossingcondition}) to hold we need
\bq \label{nocrossing}
Q_3^{(I,J)}(v_2)=Q_3^{(J,I)}(v_2) \, , \, \ka_I=\ka_J \, , \, \dot{A}_I(\lm_Iv_2)=\dot{A}_J(\lm_Jv_2) \, , \, A_I(\lm_Iv_2)=A_J(\lm_Jv_2) .
\eq
%
The conditions (\ref{nocrossing}) suggest that the supersheets can only intersect through a point at which they tangentially touch each other. It is interesting to observe that both intersecting and touching require the helical winding numbers $\ka_I$, $\ka_J$ to be the same. In both cases the requirement of a local condition (continuity) gives a constraint on the global parameters $\ka$. This constraint (along with $\dot{A}_I=\dot{A}_J$ for intersecting) can be understood as follows: for the supersheets to touch or intersect, the superthreads they consist of should, at the touching or intersection point $v=v_2$, be able to be realized as constituents of the same supersheet. Another way of realizing the constraints (\ref{touchingconditions2}), (\ref{nocrossing}) is through the angular momentum of the supersheets (\ref{multiomgexpansion}), (\ref{multiomgexpansioncomps}). The parameters $\ka_I$ generate the $\psi$, $\phi$ components of the angular momentum and $\dot{A}_I$ the $\theta$ one. Thus for the supersheets to touch or intersect regularity requires that at the point of touching they should rotate in the same manner. 

One could also argue that the cases of supersheets touching and tangentially intersecting are not essentially different since the regularity conditions for these two situations are essentially the same and one could change from one situation to the other after a piecewise relabeling of the supersheets whenever they come into contact. Such a procedure would of course change the $A(v)$ part of the profile functions of the superthreads and hence the supersheets, but that is safe to do since our solutions can be realized as a collection of different $v$ slices. Thus as far as the constraints (\ref{nocrossing}) are satisfied any pair of touching supersheets with some profile functions can be realized as a pair of tangentially intersecting supersheets with different profile functions and vice versa. Up to the change in the profile functions, both of these solutions are being described by (\ref{multisingle}), (\ref{multiinteraction}) and (\ref{multidefn}).

Another remark is that because the five-dimensional bubble equations are encoded in the ordering of the supersheets and since at a touching or intersection point the ordering becomes degenerate, the constraints (\ref{touchingconditions2}), (\ref{nocrossing}) are basically the six-dimensional conditions for the absence of Dirac strings when different supersheets touch or intersect.

\subsection{Numerics on Global Regularity Conditions} \label{numerics}
\begin{figure}[!ht] 
\begin{center}
 \subfigure[$A_1=2$ Well separated supersheets ]{\includegraphics[angle=0,
width=0.48\textwidth]{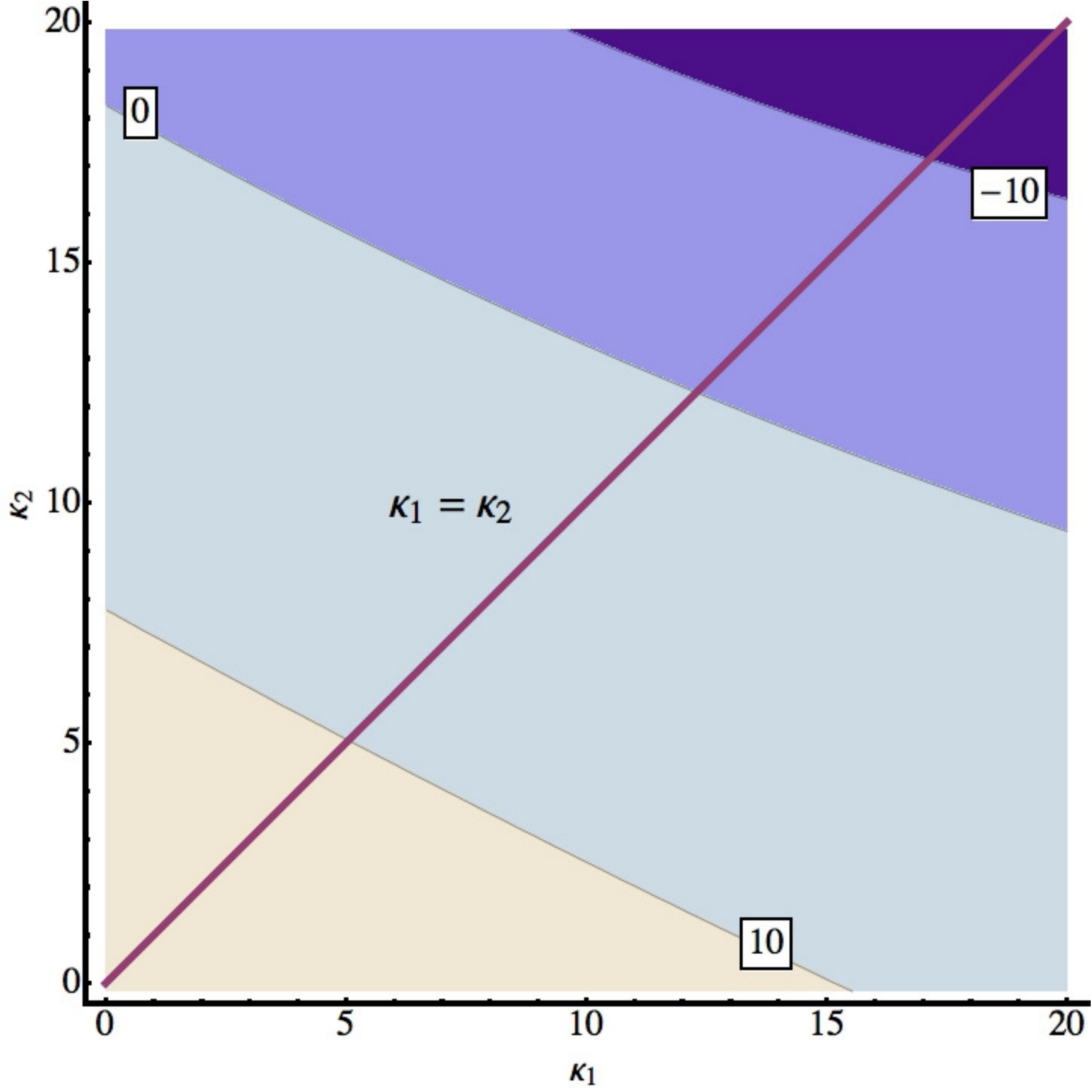} \label{fig2a}}
\subfigure[$A_1=4.9$ Supersheets close to each other]{\includegraphics[angle=0,
width=0.48\textwidth]{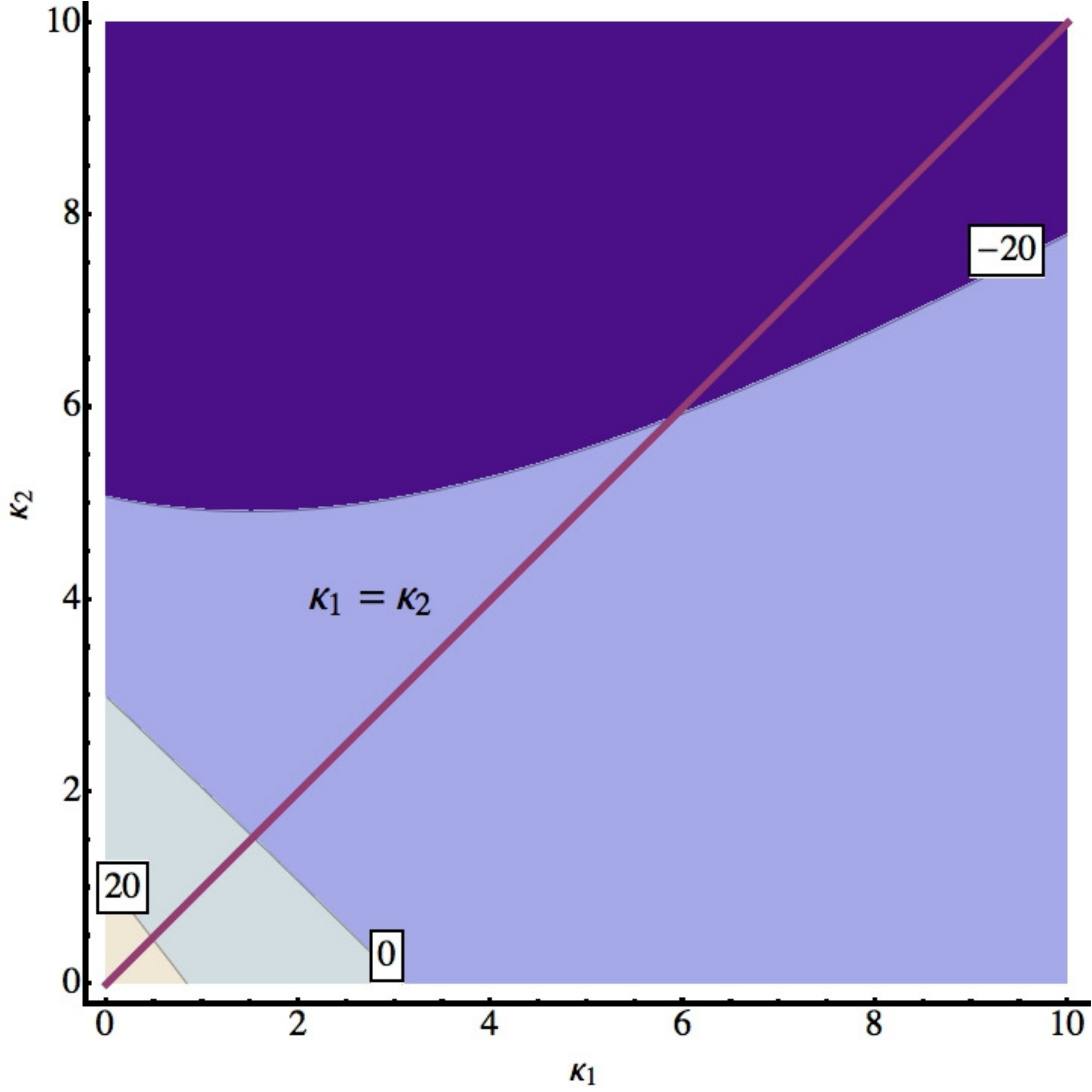}  \label{fig2b}}
\caption{\it  \small Contour plot of the $d\psi^2$ component of the left hand side of (\ref{noCTC2}) for $\theta=\frac{\pi}{2}$ shows the allowed values of $\ka_1$, $\ka_2$ for separated supersheets in the middle of the area between them. The left hand side of (\ref{noCTC2}) has to be greater or equal to zero. The allowed values of $\kappa_1$ and $\kappa_2$ reduce as the supersheets approach.}
\label{fig2}
\end{center}
\end{figure}

\begin{figure}[!ht] 
\begin{center}
\includegraphics[width=9cm]{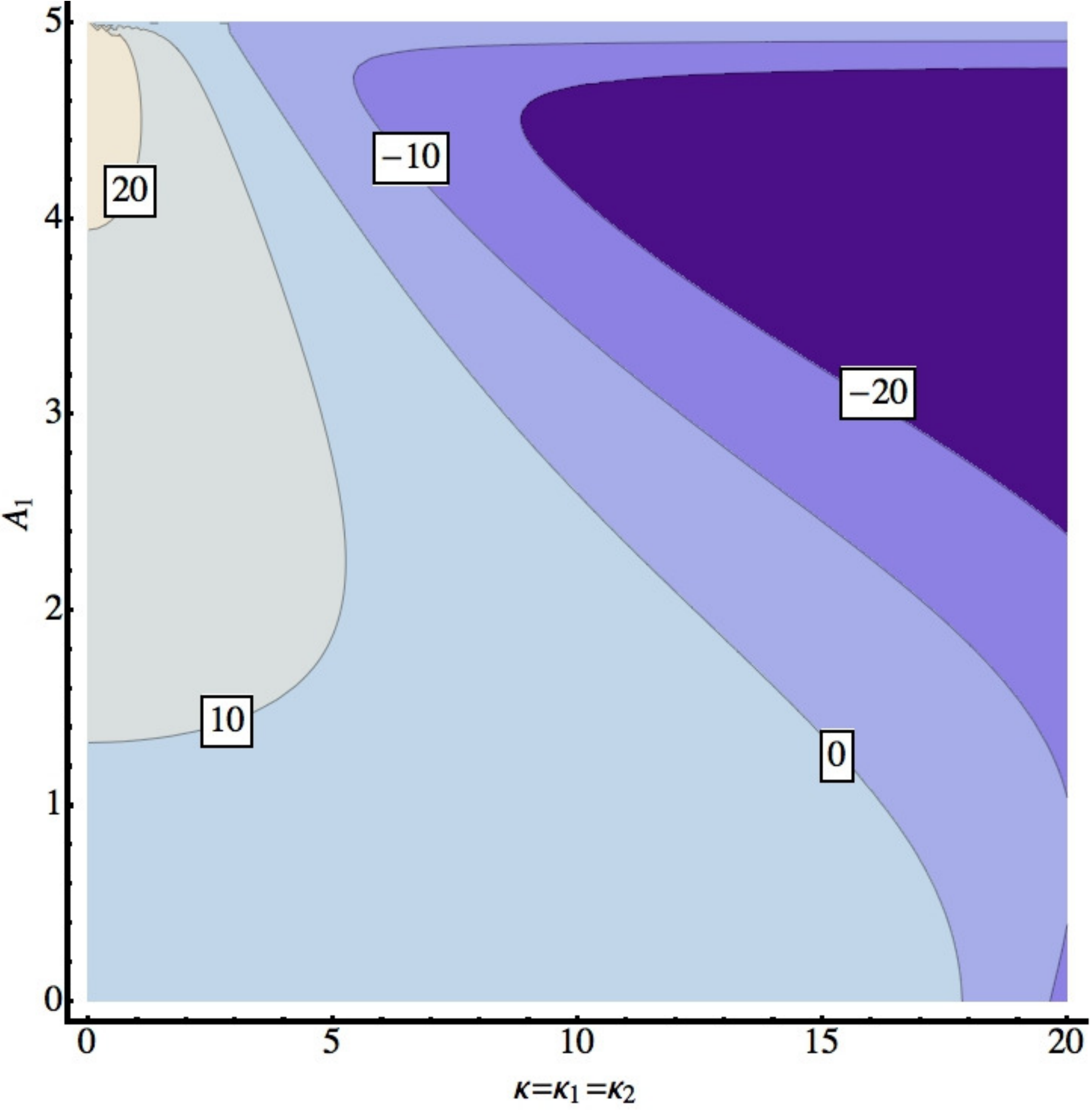}
\caption{ \small \it  Contour plot of the $d\psi^2$ component of the left hand side of (\ref{noCTC2}) for $\theta=\frac{\pi}{2}$ in the middle of the area between two supersheets with same helical mode $\ka$. The left hand side of (\ref{noCTC2}) has to be greater or equal to zero. The allowed values of $\ka$ reduce as the supersheets approach.}
\label{figure3}
\end{center}
\end{figure}

\begin{figure}[!ht]
\begin{center}
\includegraphics[width=9cm]{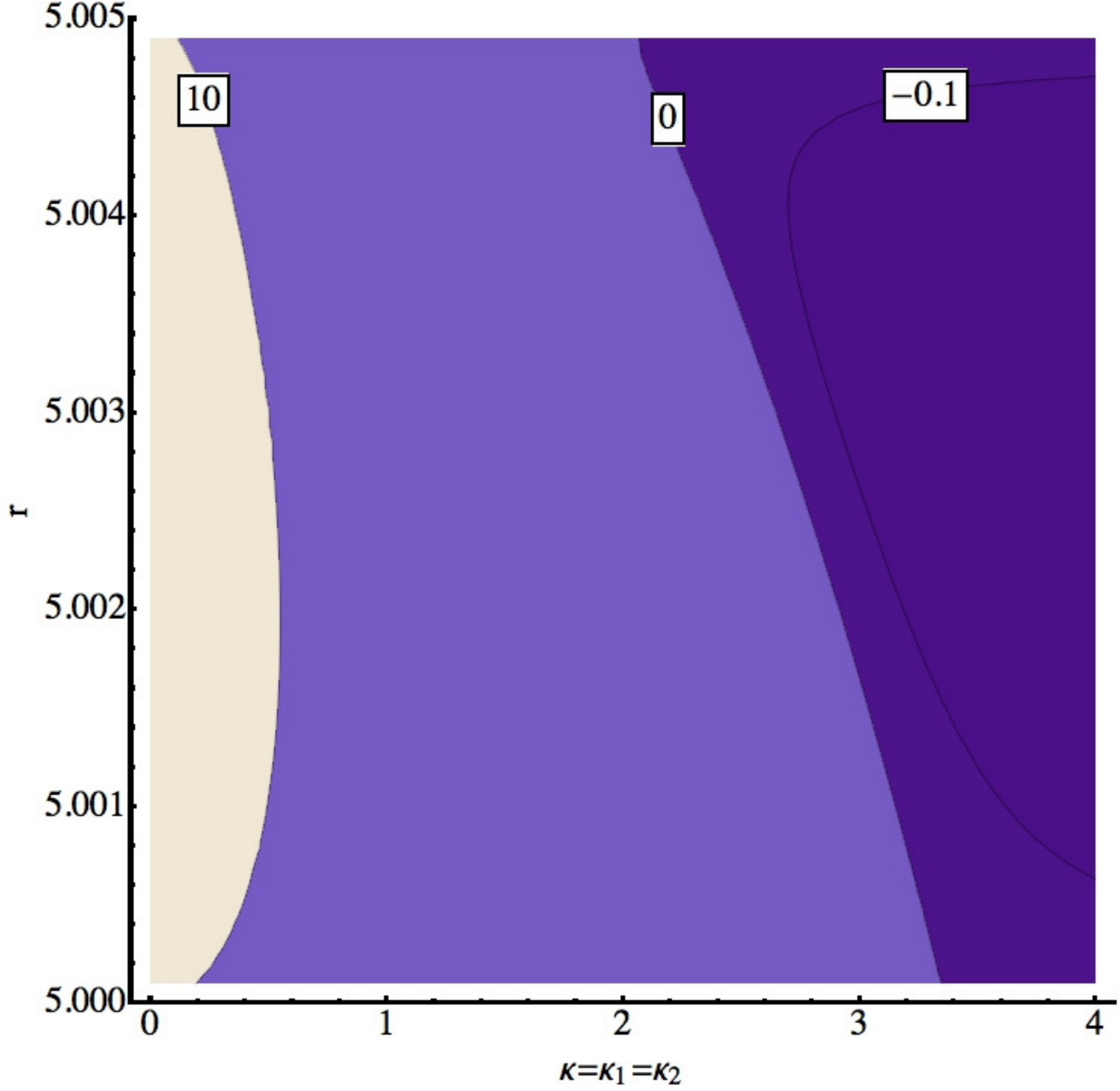}
\caption{ \small \it  For touching supersheets there are allowed values of $\ka$ for all radial distances $r$ in the area between them next to the touching point.}
 \label{fig4}
\end{center}
\end{figure}

\begin{figure}[!ht]
\begin{center}
\includegraphics[width=9cm]{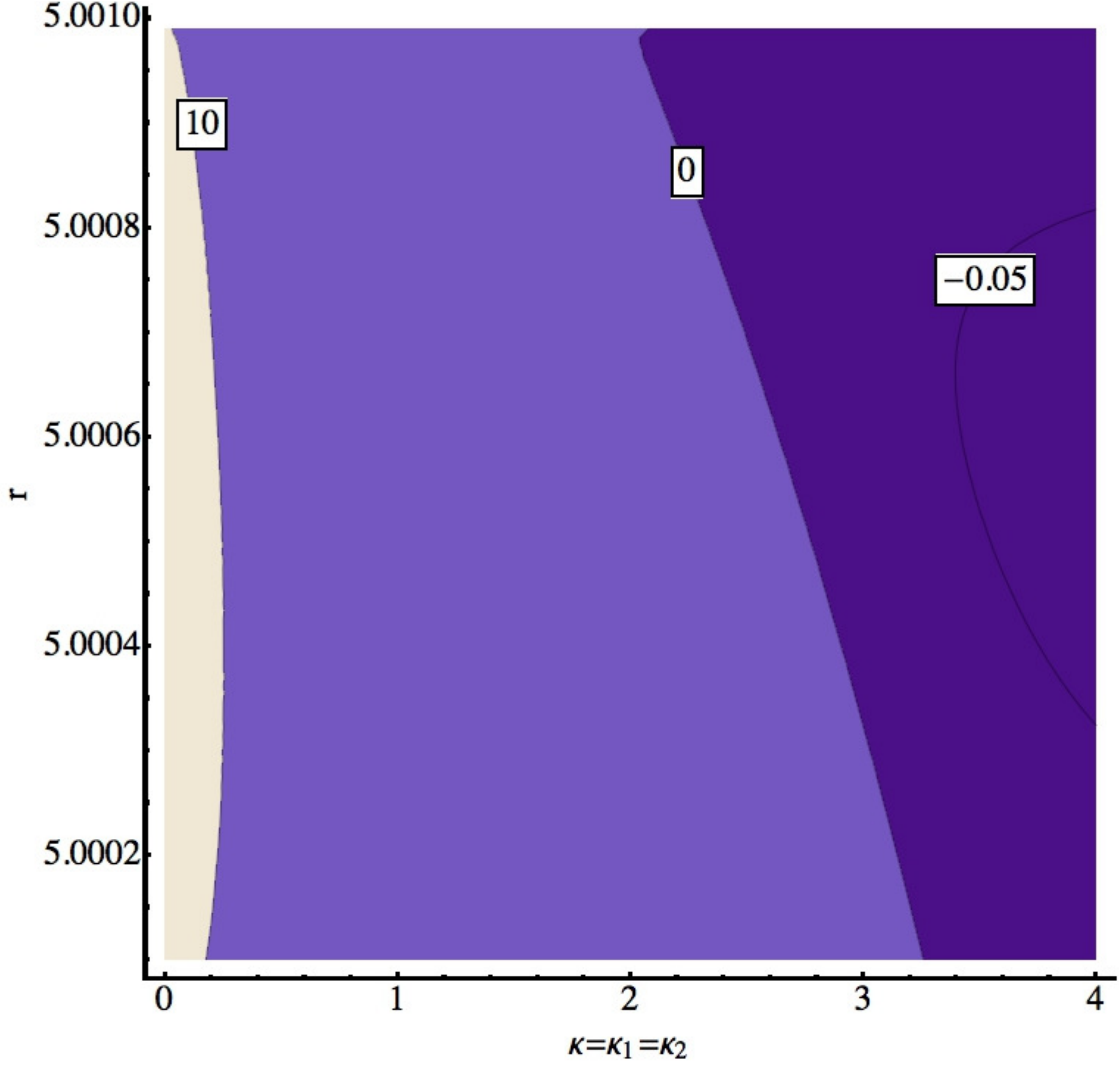}
\caption{ \small \it  For intersecting supersheets there are allowed values of $\ka$ for all radial distances $r$ in the area between them next to the intersection point.}
 \label{fig5}
\end{center}
\end{figure}

So far we have examined the case of well separated multi-supersheets as well as the situation at which they touch or intersect. In our analysis we mainly focused in the local regularity conditions that these solutions have to satisfy. However one should also check the global constraints (\ref{noCTC1}), (\ref{noCTC2}) which come from the absence of closed timelike curves. Here by considering specific examples of two concentric supersheets we perform a partial numerical analysis of these constraints and observe that at the usual areas of danger, our solutions pass the test. We will focus in the area between the supersheets and examine the phase space of solutions as the two supersheets approach or as we vary the parameters $\ka_1$, $\ka_2$. Regarding (\ref{noCTC2}) we will examine the $d\psi^2$ part of it, which matches the five dimensional global regularity condition \cite{Bena:2007kg}. In all of the cases we will have a non-corrugated supersheet of constant radius $A_1$ and we choose
\bq \label{genchoices}
Q^{(1)}_1=Q^{(1)}_2=Q^{(2)}_1=Q^{(2)}_2=1 \, , \, Q^{(1)}_3=Q^{(2)}_3=Q^{(21)}_3=50  \, , \, \theta=\frac{\pi}{2} .
\eq
For distinct or touching supersheets we will choose
\bq \label{touchingA2} 
A_2=6+\sin v ,
\eq
which means that $0<A_1\le 5$ with the supersheets touching for $A_1=5$. For $A_1<5$ we will examine the area for $v=\frac{3\pi}{2}$ where the supersheets $A_1$ and $A_2$ are closest to each other and we will choose the radial distance to be in the middle of the two supersheets $r=\frac{5+A_1}{2}$. The results are shown in figures (\ref{fig2a}), (\ref{fig2b}) and (\ref{figure3}) in terms of contour plots of (\ref{noCTC2}). In fig.(\ref{fig2a}) we have chosen $A_1=2$. This corresponds to the case where the supersheets are well separated. We see that there is enough parameter space for $\ka_1$, $\ka_2$ before closed timelike curves appear. In fig.(\ref{fig2b}) we have $A_1=4.9$ and the allowed parameter space for $\ka_1$, $\ka_2$ is significantly smaller. This is displayed better in fig.(\ref{figure3}) where we have set $\ka_1=\ka_2$ (the condition for touching) and observe how the allowed values decrease as the supersheets approach.
\goodbreak
\goodbreak
For touching supersheets we have $A_1=5$ and $\ka_1=\ka_2=\ka$. We set $v=\frac{3\pi}{2} + 0.1$ and examine the area between the two supersheets near the touching point as we vary $\ka$ and the radius $r$. The results are displayed in fig.(\ref{fig4}). Once again we see there are ample values for $\ka$ so that (\ref{noCTC2}) is satisfied.
\goodbreak
For intersecting supersheets we choose
\bq \label{crossingA2}
A_1=5 \, , \, A_2=5+\left(\sin\left(v+\frac{\pi}{2}\right)\right)^3 .
\eq
The function $A_2$ has inflection points for $v=(2n+1)\frac{\pi}{2}$, where $n \in \ZZ$. At these points $A_1=A_2$ and that's when the two supersheets intersect. As in the case of touching supersheets we are going to choose $v=\frac{3\pi}{2}+0.1$, so that we are near the intersection point, and examine the area between $A_1$ and $A_2$. As we observe from fig.(\ref{fig5}) there are allowed values of $\ka$ from (\ref{noCTC2}) and thus the solution exists.
\goodbreak

The condition (\ref{noCTC1}) is trivially satisfied in the areas we examined so far. Examining (\ref{noCTC1}) near the origin we also need (\ref{Freg}), (\ref{Q3intercondition}). Also from examining (\ref{noCTC2}) near the supersheets when $\dot{A}_I=0$ we additionally get (\ref{chargemomentum}). All of these conditions (\ref{Freg}), (\ref{Q3intercondition}), (\ref{chargemomentum}) set finite upper bounds to the values of $\ka_1$, $\ka_2$. Upper bounds to the values of $\ka_1$, $\ka_2$ have also been found in the numerical analysis  presented in this section. Thus the conditions (\ref{Freg}), (\ref{Q3intercondition}), (\ref{chargemomentum}) together with examining (\ref{noCTC2}) in the area between the supersheets will have a common set of allowed values for $\ka_1$, $\ka_2$. 

\goodbreak
\section{Suggestions on the Addition of KKM} \label{KKM}
The geometries we have studied so far are restricted to setting $\beta=0$ and hence there is no third dipole charge. The addition of $\beta$ is needed to construct not only generic black hole solutions, but also the superstratum. Here we use the results of this paper to argue about certain aspects of the solution that include $\beta$ \footnote{Geometries with all three electric and dipole magnetic charges (they have $\beta\neq 0$ but still $\dot{\beta}=0$) have been constructed in \cite{Bobev:2012af} for a K\"{a}hler base space. In contrast to five dimensions, in six dimensions the BPS conditions do not require the base space to be hyper-K\"{a}hler \cite{Bena:2011dd, Gutowski:2003rg} and indeed the more general solutions we are after may require a more general class of metrics. However it is still interesting to examine which genuinely six-dimensional geometries can be constructed with a hyper-K\"{a}hler base.}. We find that our arguments are consistent with previous analysis regarding the superstratum \cite{Bena:2011uw}.

The difficulty arises because the BPS equation \cite{Bena:2011dd} for $\beta$ is non-linear
\bq \label{Dbeta}
\mathcal{D}\beta = *_4 \mathcal{D}\beta ,
\eq
where
\bq \label{D}
D\Phi=\tilde{d}\Phi- \beta\wedge \dot{\Phi}
\eq
and $\tilde{d}$ is the exterior derivative with respect to the four-dimensional base space. When $\beta$ is $v$-independent (\ref{Dbeta}) becomes linear and reduces to the equation known from linearity in five dimensions
\bq \label{dbeta}
\tilde{d}\beta = *_4\tilde{d}\beta .
\eq

In section \ref{regularity} we observed that because of the periodicity of $v$, there should be values of $v=v_1$ such that $\dot{A}(\lm v_1)=0$. At these points our solution directly reduces to the non-corrugated supersheet which is essentially five-dimensional. Thus for any generic genuinely six-dimensional solution there are points along the compactification circle such that the solution looks five-dimensional. Consequently, we should be able to explore six-dimensional solutions by adding appropriate perturbations around the points $v=v_1$ to already known geometries from five dimensions. This should be allowed based on the superposition of corrugated and helical modes we constructed in section \ref{regularity}. 
At $v=v_1$ we should simultaneously have
\bq \label{v1conditions}
\dot{A}(v_1)=0 \, , \, \dot{\beta}(v_1)=0 ,
\eq
with $A(v)$ representing the radius of the profile in the four-dimensional base metric.Thus by expanding around $v=v_1$, $\beta$ should at least have similar linear order dependence on $v$ with $A$.\\

The superstratum, since it is a smooth geometry, is expected to be much more constrained compared to generic black geometries and one would have to add appropriate perturbations in a very precise manner. Indeed, the only known D1-D5 geometry that has all of its electromagnetic sources at a single point, is smooth in six dimensions and can be reduced to five is the supertube. The supertube carries D1, D5 electric charges and KKM dipole charge. The superstratum is expected to have all three electric (D1-D5-P) and magnetic dipole (d1-d5-KKM) charges. Thus for $v=v_1$ the superstratum should look exactly like a supertube aligned along the $v$ direction and because of the smoothness of the solution we should simultaneously have
\bq \label{stratumconditions}
\begin{split}
& \dot{A}(v_1)=0 \, , \, \dot{\beta}(v_1)=0 ,\\
& P(v_1)=0\, , \, d1(v_1)=d5(v_1)=0 .
\end{split}
\eq
The additional requirement on the charges of the object is consistent with the supersymmetry analysis in \cite{Bena:2011uw}. The D1-D5 system is being placed along the $v$ direction and is being given momentum $P$ vertically with respect to the branes. This gives a system with charges $Q_1=Q_{D1}$, $Q_2=Q_{D5}$ and angular momentum $J=P$. To generate dipole charge and momentum along $v$ one gives a tilt of angle $\al$ to the D1-D5 system with respect to the $v$ direction. Then
\bq \label{tiltedcharges}
d1=Q_{D1}\sin\al \, , d5=Q_{D5}\sin\al \, , \, P_v=P\sin\al .
\eq
Consequently, the conditions (\ref{stratumconditions}) on the charges can be simultaneously satisfied provided the tilting angle $\al$ is a function of $v$ such that
\bq \label{tiltingangle}
\al=\al(v)\, , \, \al(v_1)=n\pi ,
\eq
where $n$ is an integer. The dependence of the tilting angle with respect to $v$ is an essential part of the second supertube transition \cite{Bena:2011uw} which is in turn required to generate the superstratum.

\section{Conclusions} \label{conclusions}
Minimal ungauged $\mathcal{N}=1$ supergravity in six dimensions coupled to one anti-self-dual tensor multiplet can be reduced to five-dimensional $\mathcal{N}=2$ ungauged supergravity coupled to two vector multiplets. In recent work \cite{Bena:2011dd} the BPS equations of the six-dimensional theory were shown to admit a linear structure. This allows new classes of black hole and microstate geometries to be constructed that are part of the D1-D5 system and have $AdS_3\times S^3$ asymptotics. In general these geometries would have three electric (D1-D5-P) and three dipole magnetic charges (d1-d5-KKM).

Here we extended the results of previous work \cite{Niehoff:2012wu} to describe multi-supersheet solutions with D1-D5-P electric charges and d1-d5 magnetic dipoles, which have a non-trivial dependence on the compactification direction. The supersheet profile functions were chosen so that they are the product of a helical mode and an arbitrary corrugated mode. Then we saw that after smearing the resulting supersheet is basically the superposition of the two modes.
When multiple supersheets are present they interact in pairs and we observed that the absence of integrability conditions is encoded as a lack of symmetry in exchanging the two supersheets in the functions describing the interaction. Before smearing the integrands are symmetric in exchanging the supersheets, but while performing the integral one has to choose an order to decide which poles are within the unit circle. The asymptotic charges get contributions from the interaction terms between the supersheets, but interestingly enough these extra terms do not get affected by the non-trivial dependence along the compactification direction. Thus the contribution of the interaction to the asymptotic charges is essentially five-dimensional.

A new feature is that different supersheets can under certain conditions touch and intersect through each other without the occurrence of closed timelike curves. The local regularity conditions (originating from the absence of Dirac strings in five dimensions) require that the supersheets come in contact tangentially and also have the same helical mode windings $\ka_I$. These conditions essentially guarantee that the superthreads making the two supersheets can at the point of contact realize themselves as constituents of a single supersheet as well as that touching and intersecting supersheets are equivalent up to an appropriate relabeling of the profile functions. These features are a consequence of the fact that our solutions can be realized as a sequence of different slices of constant $v$ and it would be interesting to examine whether such features hold in more general six-dimensional solutions with $\beta\neq 0$.

Another interesting aspect of the solutions is that because the direction $v$ is periodic there are values $v_1$ of $v$ where the first derivative of the radius of the four-dimensional profile $A(v)$ vanishes and thus the solution looks five-dimensional. This, together with the superposition of corrugated and helical modes, gives hope that one will be able to add perturbative modes on already known five-dimensional solutions to generate a richer class of solutions.  We should note however that when $\beta\neq 0$ there might be a wider class of solutions such that there is no superposition between helical and corrugated modes.

All in all, this paper by considering a specific class of examples reveals some of the characteristics of six-dimensional D1-D5-P geometries and to the best of our knowledge the first example of a case with electromagnetic sources at multiple points. The expectation is that the results of our analysis will help us find more general classes of geometries, the conjectured superstratum,  construct multiple black and microstate systems as well as their non-supersymmetric analogs. We leave the exploration of these more general geometries for the future.

\section*{Acknowledgements}

I would like to thank Nick Warner for very valuable conversations and Ben Niehoff for helpful discussions during the early stages of this project. This work is supported in part by DOE grant DE-FG03-84ER-40168. OV would like to thank the USC Dana and David Dornsife College of Letters, Arts and Sciences for support through the College Doctoral Fellowship.




\end{document}